\documentclass{article}
\usepackage{amssymb}
\newcommand\maath{\mathsurround=0pt}
\newcommand{\EQM}[1]{\vcenter{\normalbaselines\maath
     \ialign{${\displaystyle ##}$\hfil&&\ ${\displaystyle ##}$\hfil\crcr
     \mathstrut\crcr\noalign{\kern-\baselineskip}
     \noalign{\smallskip}
     #1\crcr\mathstrut\crcr\noalign{\kern-\baselineskip}}}}

\newcommand{\CP}{\mathbb{CP}}

\newcommand{\Z}{\mathbb{Z}}

\begin{document}

{\bf Are Palmore's ``ignored estimates'' on the number of planar central configurations correct?}

\bigskip

\centerline{Alain Albouy, Alain.Albouy@obspm.fr}
\centerline{IMCCE, CNRS, Observatoire de Paris,}
\centerline{77, avenue Denfert-Rochereau, 75014 Paris, France}

\bigskip

{\bf Abstract.} We wish to draw attention on estimates on the number of relative equilibria in the Newtonian $n$-body problem that Julian Palmore published in 1975.

\bigskip

Julian Palmore published in \cite{Pa1} a first estimate on the number of planar central configurations with given number of bodies $n$ and given positive masses $m_1,\dots, m_n$. If these central configurations are non-degenerate, they are at least $(3n-4)(n-1)!/2$.

The word ``configuration'' is ambiguous. Here we consider two ``configurations'' which are deduced one from the other by homothety and rotation as the same configuration. But we consider that applying a reflection to a non-collinear configuration produces a distinct configuration. We also consider that the bodies are ``distinguishable'': configurations which differ only in the numbering of the bodies are nevertheless considered as distinct.

Palmore published in \cite{Pa2} a second estimate for the same number, under the same non-degeneracy hypothesis. There are at least $(n-2)!(2^{n-1}(n-2)+1)$ planar central configurations of $n$ bodies. We call this estimate the ``ignored'' Palmore estimate, because the subsequent authors on lower bounds did not even mention it (maybe because the proof is missing). Palmore also gave detailed estimates, i.e.\ a lower bound on the number of central configuration with given index. We do not know if these estimates are true. They seem however compatible with all the known results.

When we speak of degeneracy or index, we think of central configurations as critical points of a function. This function is the Newtonian potential $U=\sum_{i<j}m_im_j/r_{ij}$, where $r_{ij}$ is the distance from body $i$ to body $j$. The potential $U$ is defined on the configuration space, and should be restricted to the quadric $I=1$, where $I=\sum_{i<j}m_im_jr_{ij}^2/(m_1+\cdots+m_n)$. One should quotient out the rotations and the translations to get the good concept of configuration. The non-degeneracy can only occur at the level of the quotient space. Our convention is that a local minimum has zero index. 

1) The oldest estimate is as follows. The configuration space is $\CP_{n-2}$ with the collisions removed. And actually, when $U_0$ tends to $+\infty$, the set of points satisfying $U\le U_0$ fills up this space. This space is homotopically a ``bouquet of circles'' (see \cite{Arn}, \cite{DFN}, p.\ 324). The Poincar\'e polynomial is $(1+2t)(1+3t)\cdots(1+(n-1)t)$.

2) The first Palmore estimate uses this argument and further information. There are $n!/2$ collinear central configurations (Moulton) and their index is $n-2$ (Conley, see \cite{Pac}). So, we add to the previous estimate $n!/2-(n-1)!$ saddles of index $n-2$. But, as the Morse polynomial is obtained from the Poincar\'e polynomial by adding terms of the form $t^k+t^{k+1}$, we should add the same number of saddles of index $n-3$ (saddles of index $n-1$ do not exist, see \cite{Pa1}). In total we have at least $n!/2+2(n!/2-(n-1)!)=(3n-4)(n-1)!/2$ central configurations.

3) Another estimate was obtained by Christopher McCord in \cite{McC}. It takes into account the reflection symmetry,  and corresponds to equivariant Morse theory (see \cite{Bot}). We should compute homology and cohomology with coefficients in the field $\Z/2\Z$. There is a simple way to reach the formulas of \cite{McC}, which is inspired by the examples given in  \cite{Bot}. We do not know any rigorous argument that would justify this simplification of \cite{McC}. Instead of the Poincar\'e polynomial, we should write the fraction $(1+2t)(1+3t)\cdots(1+(n-1)t)/(1-t)$. The Morse inequalities take the usual form, but we should divide the contribution of an invariant critical point by $1-t$. Here, the invariant critical points are the configurations that are invariant by reflection, i.e., collinear. We write $(1+2t)(1+3t)\cdots(1+(n-1)t)/(1-t)=(n!/2)t^{n-2}/(1-t)+Q(t)$, where $Q(t)$ is a polynomial. If $R(t)=a_0+a_1t+\cdots$, where $2a_0$ is the number of non-collinear local minima, $2a_1$ the number of non-collinear saddles of index 1, etc., then $R(t)-Q(t)=(1+t)S(t)$, where $S(t)$ is a polynomial with non-negative integer coefficients.

4) Palmore's ``ignored estimates'' are developed in \cite{PaM}, where a cellular decomposition of the configuration space is considered. Palmore does not explain why $U$ should have the critical points corresponding to this cell decomposition. Palmore predicts $(n-1)!$ local minima, $n(n-2)(n-2)!$ saddles of index 1, etc. The only mentioned property of $U$ is the reflection symmetry, which is likely to be fully taken into account by McCord.  Concerning the local minima, McCord predicts only two of them, and it is probably impossible to predict more than two without using further properties of $U$. In the first non-trivial case, $n=4$, Palmore's estimates are easily proved by using a result by McMillan and Bartky (see \cite{McB} and \cite{Xia}): there is at least one local minimum, a convex quadrilateral, for each cyclic ordering of the four bodies. This gives, compared to the first Palmore estimate, 5 more local minima, and we should consequently also add 5 saddles of index 1. We get 34 central configurations in total, which is Palmore's bound.
The argument of  \cite{McB} and \cite{Xia} is basically that no central configuration can cross the border from convex to non-convex. This is no longer true with five bodies in a plane, as discovered independently in \cite{GLl} and in \cite{ChH}. We also see in the lists \cite{MoN} or \cite{Fer} of central configurations with equal masses that the index of the convex configurations is growing when $n$ is growing (see also \cite{Woe}).

5) Zhihong Xia gave in \cite{Xi1} the exact number of central configurations in the case $m_1\gg m_2\gg \cdots \gg m_n$. The explicit formulas corresponding to his construction do not appear in his paper, but Moeckel and Tien computed them (see \cite{MoM}, p.\ 81). Surprisingly, the recursion formulas they found are those used by Palmore to get his ``ignored estimate''. The numbers are the same, and the detailed numbers index by index are obtained by a reasonable guess: that the index of a configuration obtained by Xia's construction is the number of times a saddle was chosen as the position of the next body.

We give below the numbers corresponding to these estimates for $n=3$, $4$ and $5$. We add some examples. Examples with $n=4$ are from \cite{Pa3} and \cite{Sim}. Examples with $n=5$ are from \cite{Si2}, \cite{MoN} and \cite{Fer}. According to Carles Sim\'o, assuming non-degeneracy, the number of planar central configurations with $n=5$ and given positive masses is very likely to be always  between 294 and 450. 

I wish to thank Alain Chenciner, Joseph Fayad, Chris McCord and Rick Moeckel for their help.

$$\EQM{n=3&\hbox{Index 0 }&\hbox{Index 1 }&\hbox{Total}\cr
\hbox{bouquet}&1&2&3\cr
\hbox{first Palmore }&2&3&5\cr
\hbox{McCord }&2&3&5\cr
\hbox{Ignored Palmore }&2&3&5\cr
\hbox{All examples }&2&3&5}$$

$$\EQM{n=4&\hbox{Index 0 }&\hbox{Index 1 }&\hbox{Index 2 }&\hbox{Total}\cr
\hbox{bouquet }&1&5&6&12\cr
\hbox{first Palmore }&1&11&12&24\cr
\hbox{McCord }&2&12&12&26\cr
\hbox{Ignored Palmore }&6&16&12&34\cr
\hbox{Xia's case }&6&16&12&34\cr
\hbox{3 equal, 1 small }&8&18&12&38\cr
\hbox{Equal masses }&6&24&20&50
}$$

$$\EQM{n=5&\hbox{Index 0 }&\hbox{Index 1 }&\hbox{Index 2 }&\hbox{Index 3 }&\hbox{Total}\cr
\hbox{bouquet }&1&9&26&24&60\cr
\hbox{first Palmore }&1&9&62&60&132\cr
\hbox{McCord }&2&20&72&60&154\cr
\hbox{Ignored Palmore }&24&90&120&60&294\cr
\hbox{Xia's case }&24&90&120&60&294\cr
\hbox{Equal masses }&54&120&120&60&354\cr
\hbox{4 equal, 1 small }&30&120&192&108&450
}$$


\begin{thebibliography}{99}

\parskip0pt

\itemsep0pt

\bibitem{Arn} V.I. Arnold, {\it The cohomology ring of dyed braids}, Mat.\ Zametki 5 (1969), pp.\ 227--231

\bibitem{Bot} R. Bott,  {\it Lectures on Morse theory, old and new}, Bull.\ Amer.\ Mat.\ Soc.\  7 (1982), pp.\ 331--358



\bibitem{ChH} K.-C. Chen, J.-S. Hsiao, {\it Convex central configurations of the $n$-body problem which are not strictly convex}, Journal of Dynamics and Differential Equations 24 (2012), pp.\ 119--128

\bibitem{DFN} B.A. Dubrovin, A. T. Fomenko, S. P. Novikov, {\it Modern geometry - methods and applications: Introduction to homology theory}, Springer-Verlag, Berlin, 1990

\bibitem{Fer} D.L. Ferrario,  {\it Central configurations, symmetries and fixed points}, arXiv preprint math/0204198 (2002), 47 pp.

\bibitem{GLl} M. Gidea, J. Llibre, {\it Symmetric planar central configurations of five bodies: Euler plus two}, Celestial Mechanics and Dynamical Astronomy 106 (2010), pp.\ 89--107

\bibitem{McC} C.K. McCord, {\it Planar central configuration estimates in the n-body problem}, Ergodic Theory and Dynamical Systems 16 (1996), pp.\ 1059--1070

\bibitem{MoM} R. Moeckel, {\it Celestial Mechanics --- especially central configurations}, hand-written lecture notes (1994), 124 pp.

\bibitem{MoN} R. Moeckel, {\it Some Relative Equilibria of N Equal Masses}, preprint (1989)
Russian translation in: Otnositelnye ravnovesita. Periodicheskie resheniya (Relative Equilibria. Periodic Solutions), Izhevsk: Inst.\ Kompyut.\ Issled., 2006, pp.\ 83--98

\bibitem{McB} W.D. MacMillan, W. Bartky, {\it Permanent configurations in the problem of four bodies}, Trans.\ Amer.\ Math.\ Soc.\ 34 (1932), pp.\ 838--875

\bibitem{Pac} F. Pacella, {\it Central configurations of the N-body problem via
equivariant Morse theory},  Archive for Rat.\ Mech.\ and Anal.\ 97 (1987), pp.\ 59--74

\bibitem{Pa1} J.I. Palmore, {\it Classifying relative equilibria. I}, Bull.\ Amer.\ Mat.\ Soc.\ 79 (1973), pp.\ 904--908

\bibitem{Pa2} J.I. Palmore, {\it Classifying relative equilibria. II}, Bull.\ Amer.\ Mat.\ Soc.\ 81 (1975), pp.\ 489--491

\bibitem{Pa3} J.I. Palmore, {\it Classifying relative equilibria. III}, Letters in Mathematical Physics 1 (1975), pp.\ 71--73

\bibitem{PaM} J.I. Palmore,  {\it  Minimally classifying relative equilibria}, Letters in Mathematical Physics 1 (1977), pp.\ 395--399

\bibitem{Sim} C. Sim\'o, {\it Relative equilibrium solutions in the four body problem}, Celestial Mechanics 18 (1978), pp.\ 165--184

\bibitem{Si2} C. Sim\'o, {\it Dynamical systems, numerical experiments and super-computing}, Memorias de la Real Academia de Ciencias y Artes de Barcelona 987 (2003), pp.\ 3--36

\bibitem{Woe} K.D.P. Woerner, {\it The $N$-gon is not a local minimum of $U^2I$ for $N\geq 7$}, Celestial Mechanics and Dynamical Astronomy 49 (1990), pp.\ 413--421

\bibitem{Xia} Z. Xia, {\it Convex central configurations for the $n$-body problem}, J.\ Differential Equations 200 (2004), pp.\ 185--190

\bibitem{Xi1}  Z. Xia, {\it Central configurations with many small masses}, J.\ Differential Equations 91 (1991), pp.\ 168--179

\end{thebibliography}
\end{document}